\documentclass[12pt]{article}

\usepackage{epsfig}
\usepackage{subfigure}

\usepackage{amssymb}
\usepackage{amsthm}
\usepackage[cmex10]{amsmath}
\usepackage{hyperref}

\newtheorem{theorem}{Theorem}[section]

\newtheorem{proposition}{Proposition}[section]

\newtheorem{remark}{Remark}[section]

\def\A{{\mathcal{A}}}

\def\e{\mathbf{e}}

\def\u{\mathbf{u}}
\def\P{\mathbf{P}}
\def\0{\mathbf{0}}
\def\1{\mathbf{1}}

\def\E{\mathbb{E}}
\def\K{\mathbf{K}}

\def\L{\mathbf{L}}
\def\R{{\mathbb{R}}}
\newcommand{\argmax}{\mathop{\rm argmax}\limits}
\newcommand{\argmin}{\mathop{\rm argmin}\limits}
\def\proof {{Proof.} }
\def\endproof{\hfill $\Box$ \vskip .3cm}

\begin{document}

\title{Time Consistent Behavioral Portfolio Policy for Dynamic Mean-Variance Formulation\thanks{This work was partially supported by Research Grants Council of Hong Kong under grants 414207, 519913 and 15209614, by National Natural Science Foundation of China under grant 71201094 and Shanghai Pujiang Program 12PJC051. The third author is also grateful to the support from the Patrick Huen Wing Ming Chair Professorship of Systems Engineering \& Engineering Management.}}

\author{Xiangyu Cui \thanks{School of Statistics and Management, Shanghai University of Finance and Economics, and Key Laboratory of Mathematical Economics (SUFE),  Ministry of Education, China}, \quad
Xun Li\thanks{Department of Applied Mathematics, The Hong Kong Polytechnic University, Hong Kong}, \quad
Duan Li\thanks{Corresponding author. Department of Systems Engineering and Engineering Management, The Chinese University of Hong Kong, Hong Kong},\quad
and \quad Yun Shi \thanks{School of Management, Shanghai University, Shanghai, China}}

\date{}
\maketitle

\begin{abstract}
When one considers an optimal portfolio policy under a mean-risk
formulation, it is essential to correctly model investors' risk
aversion which may be time variant, or even state-dependent. In this
paper, we propose a behavioral risk aversion model, in which risk
aversion is a piecewise linear function of the current wealth level
with a reference point at the discounted investment target. Due to
the time inconsistency of the resulting multi-period mean-variance
model with adaptive risk aversion, we investigate the
time consistent behavioral portfolio policy by solving a nested
mean-variance game formulation. We derive a semi-analytical time
consistent behavioral portfolio policy which takes a piecewise
linear feedback form of the current wealth level with respect to the
discounted investment target. Finally, we extend our results on time
consistent behavioral portfolio selection to dynamic mean-variance
formulation with a cone constraint.

\end{abstract}

\noindent
{\bf Key Words:}
risk aversion, mean-variance formulation,  time consistent behavioral portfolio policy.

\section{Introduction}

According to the classical investment doctrine in \cite{Markowitz}, an investor of a mean-variance type needs to strike a balance between maximizing the expected value of the terminal wealth, $\E[X_1|X_0]$, and minimizing the investment risk measured by the variance of the
terminal wealth,  $\mbox{Var}(X_1|X_0)$, by solving the following mean-variance formulation,
\begin{equation*}
\begin{array}{rrl}
(MV(\gamma)):~~ & \min & \mbox{Var}(X_1|X_0)-\gamma\E[X_1|X_0],
\end{array}
\end{equation*}
where $X_0$ is the initial wealth level, $X_1$ is the terminal wealth at the end of the (first) time period and $\gamma\geq 0$ is the trade-off parameter between the two conflicting objectives. We call $\gamma$ the risk aversion parameter, which represents the risk aversion attitude of the investor. The larger the value of $\gamma$, the less the risk aversion of the investor. Mathematically, $(MV(\gamma))$ is equivalent to the following formulation,
\begin{equation*}
\begin{array}{rrl}
(MV(\omega)):~~ & \max & \E[X_1|X_0]-\omega\mbox{Var}(X_1|X_0),
\end{array}
\end{equation*}
with the risk aversion parameter $\omega=1/\gamma$.

In a dynamic investment environment, the risk aversion attitude of a mean-variance investor may change from time to time, or could be even state-dependent (i.e., dependent on the investor's current wealth level $X_t$ realized at time $t$). \cite{Bjork:2014} and \cite{Wu2013} proposed, respectively, in continuous-time and multi-period settings, that the risk aversion parameter $\omega$ takes the following simple form of the current wealth level $X_t$,
\begin{align*}
\omega(X_t)=\frac{\omega}{X_t},\quad (\omega\geq 0).
\end{align*}
Due to the positiveness of the wealth process $X_t$ in the continuous-time setting, $\omega(X_t)$ proposed by \cite{Bjork:2014} is always nonnegative and is a decreasing function of the current wealth level. Applying the same model to a multi-period setting as proposed in \cite{Wu2013}, however, could encounter some problem, as there is no guarantee for the positiveness of the wealth process in a discrete-time setting. When the wealth level is negative, $\omega(X_t)$ becomes negative, which leads to an irrationality of the investor to maximize both the expected value and the variance of the terminal wealth, resulting in an infinite position on the riskiest asset (See Theorem 7 in  \cite{Wu2013}).

In a continuous time setting, \cite{HuJinZhou2012} introduced the risk aversion parameter $\gamma$ as a linear function of the current wealth level $X_t$,
\begin{align*}
\gamma(X_t)=\mu_1X_t+\mu_2,\quad (\mu_1\geq 0).
\end{align*}
When the wealth level is less than $-\mu_2/\mu_1$, $\gamma(X_t)$ becomes negative. It also leads to an irrationality of the investor, which contradicts the original interests of the investor of a mean-variance type.

In this paper, we propose a behavioral risk aversion model as follows,
\begin{align}\label{eq:piecewise-risk-aversion}
\gamma_t(X_t)=\left\{\begin{array}{rl}
\gamma_t^+ (X_t-\rho_t^{-1}W),  & \mbox{if } X_t\geq \rho_t^{-1}W, \\
-\gamma_t^- (X_t-\rho_t^{-1}W), & \mbox{if } X_t< \rho_t^{-1}W,
\end{array}\right.
\end{align}
where $W$ is the investment target set by the investor at time $0$, $\rho_t^{-1}$ is the risk-free discount factor from the current time $t$ to the terminal time $T$, and $\gamma_t^+\geq  0$ and $\gamma_t^-\geq 0$ are $t$-dependent  risk aversion coefficients for the ranges of $X_t$ on the right and left sides of $\rho_t^{-1}W$, respectively. Basically, we consider a piecewise linear state-dependent risk aversion function in our behavioral risk aversion model.

This proposed behavioral risk aversion model is pretty flexible in incorporating the behavioral pattern of a mean-variance investor. If the current wealth level is the same as the discounted investment target, the investor becomes fully risk averse and thus invests only in the risk-free asset. If the current wealth level is larger than the discounted investment target, the investor may consider the surplus over the discounted target level as house money and the larger the surplus the less the risk aversion. If the current wealth level is less than the discounted target level, the investor may intend to break-even and the larger the shortage under the discounted target, the stronger the desire to break-even (the less the risk aversion). The magnitude of $\gamma_t^+$ (or $\gamma_t^-$) represents the risk aversion reduction with respect to one unit increase of the surplus (or the shortage). Apparently, different mean-variance investors may have different choices of $\gamma_t^+$ and $\gamma_t^-$. For example, an investor, who is eager for breaking-even when facing shortage and feels less sensitive with the levels of surplus, may set $\gamma_t^->\gamma_t^+$. Although we use the same terms of ``house money'' and ``break-even'' as in behavioral finance, their meanings are slightly different. In behavioral finance, the house money effect describes the behavior that people take greater risk following prior gains, while the break-even effect describes the behavior that people take greater risk following prior losses (see, for examples, \cite{Staw1976}, \cite{ThalerJohnson1990}, \cite{WeberZuchel2005}). While no risk attitude is assumed for investors in their study, all investors are assumed to be risk-averse under the dynamic mean-variance framework discussed in this paper. Nevertheless, the key concepts behind the house money and break-even effects are the same for both our study and the literature in behavior finance: investors become less risk averse when experiencing either larger gains or larger losses.

The main challenge of solving multi-period mean-variance portfolio selection problem with the proposed behavioral risk aversion parameter is the time inconsistency of the problem. To see this, let us consider the simple problem with constant risk aversion parameter. At time $0$, the investor faces the following global mean-variance portfolio selection problem over the entire investment time horizon,
\begin{equation*}
\begin{array}{rrl}
(MV_0(\gamma)):~~ & \min & \mbox{Var}(X_T|X_0)-\gamma\E[X_T|X_0],
\end{array}
\end{equation*}
whose pre-committed optimal mean-variance policy is derived by \cite{LN} and given as follows,
\begin{align*}
\u_j^* = - \E^{-1}[\P_j\P_j']\E[\P_j]s_j(X_j - \lambda_0\rho_j^{-1}), \quad j = 0,1, \cdots, T-1,
\end{align*}
where $\P_j$ is the vector of excess return rates of risky assets, $X_j$ is the wealth level at time $j$ and
\[
\lambda_0 = \rho_0 X_0 +\frac{\gamma}{2}\frac{1}{\prod_{k=0}^{T-1}(1-\E[\P_k']\E^{-1}[\P_k\P_k']\E[\P_k])}.
\]
However, for $t>0$, the investor may reconsider the mean-variance portfolio selection problem for a truncated time horizon from $t$ to $T$,
\begin{equation*}
\begin{array}{rrl}
(MV_t(\gamma)):~~ & \min & \mbox{Var}(X_T|X_t)-\gamma\E[X_T|X_t],
\end{array}
\end{equation*}
whose local optimal mean-variance policy is given by
\begin{align*}\label{eq:pi-t}
\bar{\u}_j = - \E^{-1}[\P_j\P_j']\E[\P_j]s_j(X_j - \lambda_t\rho_j^{-1}), \quad j = t,t+1, \cdots, T-1,
\end{align*}
where
\[
\lambda_t = X_t\rho_t +\frac{\gamma}{2}\frac{1}{\prod_{k=t}^{T-1}(1-\E[\P_k']\E^{-1}[\P_k\P_k']\E[\P_k])}.
\]
Since $\lambda_0 \neq \lambda_t, ~~ t = 1,2, \cdots, T-1$, this leads to
$\u_j^* \neq \bar \u_j, ~~ j = t,t+1,\cdots,T-1$, i.e., the local optimal policy is different from the pre-committed optimal policy.  This interesting phenomenon is called time inconsistency (see \cite{BasakChabakauri}, \cite{CuiLiWangZhu2012}, \cite{WangForsyth2011}). In the language of dynamic programming, the {\it Bellman's principle of optimality} is not applicable to this model formulation, as the global and local objectives are not consistent (See \cite{Artzner2007}, \cite{CuiLiWangZhu2012}). In the fields of dynamic risk measures and dynamic risk management, time consistency is considered to be a basic requirement (see \cite{Rosazza2006}, \cite{BodaFilar2006},
\cite{Artzner2007} and \cite{JobertRogers2008}).

In fact, there exists a unique trade-off $\gamma(X_t)$ which depends on the wealth $X_t$, termed as the trade-off induced by the pre-committed optimal policy, such that the optimal mean-variance policy of the truncated-time horizon problem, $(MV_t(\gamma(X_t)))$, is the same as the pre-committed optimal policy, i.e., the pre-committed optimal policy of $(MV_0(\gamma(X_0)))$ could become a time consistent policy of $(MV_t(\gamma(X_t)))$ when all the trade-offs are set as the induced trade-offs (see \cite{CuiLiWangZhu2012}). Furthermore, \cite{CuiLiWangZhu2012} showed that the trade-off induced by the pre-committed optimal policy is a linear function in terms of the current wealth level $X_t$, the initial wealth level $X_0$ and the initial risk aversion parameter $\gamma_0(X_0)$.  Thus, the induced trade-off may become negative over a finite time investment horizon, which implies that investors may take irrational actions. This actually reveals that a linear trade-off is a hidden reason behind time inconsistency. Thus, to better the performance of the dynamic mean-variance formulation, the setting of trade-off parameter should go beyond the class of linear functions.

\cite{Strotz1955} suggested two possible actions to overcome time inconsistency: (1) ``He may try to pre commit his future activities either irrevocably or  by contriving a penalty for his future self if he should misbehave", which is termed as the {\it strategy of precommitment}; and (2) ``He may resign himself to the fact of inter temporal conflict and decide that his `optimal' plan at any date is a will-o'-the-wisp which cannot be attained, and learn to select the present action which will be best in the light of future disobedience", which is termed the {\it strategy of consistent planning}. Strategy of consistent planning is also called {\it time consistent policy} in the literature. For a dynamic mean-variance model, \cite{BasakChabakauri} reformulated it as an interpersonal game model where the investor optimally chooses the policy at any time $t$, on the premise that he has already decided his time consistent policies in the future. More specifically, in a framework of time consistency, the investor faces the following nested portfolio selection problem,
\begin{equation*}
\begin{array}{rrl}
(NMV_0(\gamma)):~~~ & \displaystyle\min_{u_t} & \mbox{Var}(X_T|X_t)-\gamma\E[X_T|X_t], \\[2mm]
& \mbox{\rm s.t.}  & u_j \mbox{ solves } (NMV_j(\gamma)),~~ t\leq j\leq T,
\end{array}
\end{equation*}
with the terminal period problem given as
\begin{equation*}
\begin{array}{rrl}
(NMV_{T-1}(\gamma)):~& \displaystyle\min_{u_{T-1}} & \mbox{Var}(X_T|X_{T-1})-\gamma\E[X_T|X_{T-1}].
\end{array}
\end{equation*}
The time consistent policy is then the equilibrium solution of the above nested problem, which can be derived by a backward induction.  \cite{Bjork:2014}, \cite{HuJinZhou2012} and \cite{Wu2013} extended the results in \cite{BasakChabakauri} by considering different state-dependent risk aversion mentioned before in this section. For a general class of continuous-time mean-field linear-quadratic control problems, please refer to \cite{Yong2013}. In the original setting of dynamic mean-variance portfolio selection with constant risk aversion, the time inconsistency is caused by the appearance of the variance of the terminal wealth in the objective function, which does not satisfy the smoothing property. Our model in this study that adopts a more realistic time-varying and wealth dependent risk aversion further complicates the extent of time inconsistency, which forces us to consider time consistent policies in this paper.

In this paper, we focus on studying time consistent behavioral portfolio policies under the proposed behavioral risk aversion model. The remaining parts of this paper are organized as follows: In Section 2, we provide the basic market setting and formulate the nested mean-variance portfolio selection problem. We derive in Section 3 the semi-analytical time consistent behavioral portfolio policy, which takes a piecewise linear feedback form of the surplus or the shortage with respect to the discounted wealth target. In Section 4, we extend our main results to cone constrained markets. After we offer in Section 5 numerical analysis to show the trading patterns of investors with different risk aversion coefficients, we conclude the paper in Section 6.

\section{Market Setting and Problem Formulation}

We consider an arbitrage-free capital market of $T$-time periods, which  consists of one
risk-free asset with deterministic rate of return and $n$ risky assets with random rates of return. An investor with an initial wealth $X_0$ joins the market at time 0 and allocates wealth among the
risk-free asset and $n$ risky assets at time $0$
and the beginning of each of the
following $\left( T-1\right)$ consecutive periods. The deterministic rate of return of the risk-free asset at time period $t$ is denoted by
$s_t ~(>1)$ and the random rates of return of the risky assets at time period $t$ are denoted by the vector $\e_t = [e_t^1,\cdots,e_t^n]'$, where $e_t^i$ is the random rate of return of asset $i$ at time period $t$ and $A'$ denotes the transpose operation of matrix or vector $A$. It is assumed
that $\e_t$, $t$ = 0, 1, $\cdots$, $T$ $-$ 1, are
statistically independent, absolutely integrable continuous random vectors, whose finite first and second moments, $\mathbb{E}[\e_t]$ and $\mathbb{E}[\e_t\e_t']$, are known for every $t$ and whose covariance matrixes
$\mbox{Cov}(\e_t) =\mathbb{E}[\e_t\e_t']-\mathbb{E}[\e_t]\mathbb{E}[\e_t']$, $t$ = 0, 1, $\cdots$, $T$ $-$ 1, are positive definite\footnote{Our main results can be readily extended to situations where random vectors $\e_t$, $t$ = 0, 1, $\cdots$, $T$ $-$ 1, are correlated. This extension can be achieved based on the concept of the so-called opportunity-neutral measure introduced by \cite{CernyKallsen2009}. }. All of the random vectors are defined in a filtered probability space $(\Omega,\mathcal{F}_T,\{\mathcal{F}_t\},P)$, where $\mathcal{F}_t=\sigma(\e_0,\e_1,\cdots,e_{t-1})$ and $\mathcal{F}_0$ is the trivial $\sigma$-algebra over $\Omega$.

Let $X_t$ be the
wealth of the investor at the beginning of period $t$, and $u_t^i$,
$i = 1, 2, \cdots, n$, be the amount invested in the $i$-th risky
asset at period $t$. Then, $X_t-\sum_{i=1}^n u_t^i$ is the
amount invested in the risk-free asset at period $t$. Thus, the
wealth at the beginning of period $t+1$ is given as
\begin{align*}
X_{t+1}=s_t\Big(X_t-\sum_{i=1}^n u_t^i\Big)+\e_t'\u_t=s_tX_t+\P_t'\u_t,
\end{align*}
where $$\P_t=[P_t^1,P_t^2,\cdots,P_t^n]'=[(e_t^1-s_t),(e_t^2-s_t),\cdots,(e_t^n-s_t)]'$$ is the vector of the excess rates of return and  $\u_t=[u_t^1,u_t^2,\cdots, u_t^{n}]'$ is the portfolio policy. We confine all admissible investment policies to be
$\mathcal{F}_t$-measurable Markov control, whose realizations are in $\R^n$. Then, $\P_t$ and $\u_t$ are independent, the controlled wealth process $\{X_t\}$ is an adapted Markovian process and
$\mathcal{F}_t=\sigma(X_t)$.


An investor of mean-variance type considers the following portfolio decision problem at the beginning of period $t$,
\begin{align}
\nonumber (MV_t(\gamma_t(X_t))) \quad\quad \min &~~\mbox{\rm Var}_t(X_T)- \gamma_t(X_t) \E_t[X_T], \\
\label{prob:MV_t} \mbox{s.t.} &~~ X_{j+1} = s_jX_j+\P_j'\u_j, \quad  j=t,t+1,\cdots,T-1,
\end{align}
where $\mbox{\rm Var}_t(X_T)=\mbox{Var}(X_T|X_t)$, $\E_t[X_T]=\E[X_T|X_t]$, $\rho_t^{-1}=\prod_{j=t}^{T-1}s_j^{-1}$ is the risk-free
discount factor with $\rho_T^{-1}=1$ and $\gamma_t(X_t)$ is given by \begin{align*}
\gamma_t(X_t)=\left\{\begin{array}{rl}
\gamma_t^+ (X_t-\rho_t^{-1}W),  & \mbox{if } X_t\geq \rho_t^{-1}W, \\
-\gamma_t^- (X_t-\rho_t^{-1}W), & \mbox{if } X_t< \rho_t^{-1}W.
\end{array}\right.
\end{align*}

Due to the time inconsistency of $(MV_t(\gamma_t(X_t)))$, we aim to derive the time consistent behavioral portfolio policy. More specifically,
similar to the approach in \cite{BasakChabakauri}, we formulate the multi-period mean-variance model into an interpersonal game model in which the investor optimally chooses the policy at any time $t$, on the premise that he has already decided his time consistent policy in the future. Then the time consistent behavioral portfolio policy (or time consistent policy in short) is the optimizer of the following nested mean-variance problem $(NMV)$,
\begin{align}
\nonumber (NMV_t(\gamma_t(X_t))) \quad\quad \min_{\u_t} &~~\mbox{\rm Var}_t(X_T)- \gamma_t(X_t) \E_t[X_T], \\
\label{prob:MV-tc} \mbox{s.t.} &~~ X_{t+1} = s_tX_t+\P_t'\u_t,\\
\nonumber &~~ X_{j+1} = s_jX_j+\P_j'\u_j^{TC}, \quad j=t+1,\cdots,T-1,\\
\nonumber &~~ \u_j^{TC} \mbox{ solves } (MV_j(\gamma_j(X_j))), \quad j=t+1,\cdots,T-1,
\end{align}
with terminal period problem given as
\begin{align}
\nonumber (NMV_{T-1}(\gamma_{T-1}(X_{T-1}))) \quad\quad \min_{\u_{T-1}} &~~\mbox{\rm Var}_{T-1}(X_T)- \gamma_{T-1}(X_{T-1}) \E_{T-1}[X_T] , \\
\label{MV} \mbox{s.t.} &~~ X_t = s_{T-1}X_{T-1}+\P_{T-1}'\u_{T-1},
\end{align}
which can be solved by a backward induction. Since the stage-trade off $\gamma_t(X_t)$ reflects certain behavioral pattern of an investor in terms of his wealth level, we call the optimal policy to $(NMV)$ time consistent behavioral portfolio policy.

\section{Semi-analytical Time Consistent Policy}
In this section, we derive the semi-analytical time consistent behavioral portfolio policy. Before presenting our main results, we define the following two deterministic continuous functions, $F_t^-(\K)$ and $F_t^+(\K)$, on $\mathbb{R}^n$ for $t=0,1,\cdots,T-1$,
\begin{align*}
F_t^+(\K)=&\rho_{t+1}^2 \K' (\E_t[\P_t\P_t']-\E_t[\P_t']\E_t[\P_t])\K \\
&+\E_t\left[(2\rho_{t+1}a_{t+1}^+ + b_{t+1}^+)(s_t+\P_t'\K)^21_{\{s_t+\P_t'\K\geq 0\}}\right] \\
&+\E_t\left[(2\rho_{t+1}a_{t+1}^- + b_{t+1}^-)(s_t+\P_t'\K)^21_{\{s_t+\P_t'\K< 0\}}\right] \\
&-\left(\E_t\left[a_{t+1}^+(s_t+\P_t'\K)1_{\{s_t+\P_t'\K\geq 0\}}\right] +\E_t\left[a_{t+1}^-(s_t+\P_t'\K)1_{\{s_t+\P_t'\K< 0\}}\right]\right)^2 \\
&-2\rho_{t+1} \E_t\left[a_{t+1}^+(s_t+\P_t'\K)1_{\{s_t+\P_t'\K\geq 0\}}\right] (s_t+\E_t[\P_t']\K) \\
&-2\rho_{t+1}\E_t\left[a_{t+1}^-(s_t+\P_t'\K)1_{\{s_t+\P_t'\K< 0\}}\right](s_t+\E_t[\P_t']\K) \\
&-\gamma_t^+\left(\E_t[a_{t+1}^+(s_t+\P_t'\K)1_{\{s_t+\P_t'\K\geq 0\}}]+\E_t[a_{t+1}^-(s_t+\P_t'\K)1_{\{s_t+\P_t'\K< 0\}}]\right) \\
&-\rho_{t+1}\gamma_t^+(s_t+\E_t[\P_t']\K), \\
F_t^-(\K)=&\rho_{t+1}^2 \K' (\E_t[\P_t\P_t']-\E_t[\P_t']\E_t[\P_t])\K \\
&+\E_t\left[(2\rho_{t+1}a_{t+1}^+ + b_{t+1}^+)(s_t+\P_t'\K)^21_{\{s_t+\P_t'\K\leq 0\}}\right] \\
&+\E_t\left[(2\rho_{t+1}a_{t+1}^- + b_{t+1}^-)(s_t+\P_t'\K)^21_{\{s_t+\P_t'\K> 0\}}\right] \\
&-\left(\E_t\left[a_{t+1}^+(s_t+\P_t'\K)1_{\{s_t+\P_t'\K\leq 0\}}\right] +\E_t\left[a_{t+1}^-(s_t+\P_t'\K)1_{\{s_t+\P_t'\K> 0\}}\right]\right)^2 \\
&-2\rho_{t+1} \E_t\left[a_{t+1}^+(s_t+\P_t'\K)1_{\{s_t+\P_t'\K\leq 0\}}\right] (s_t+\E_t[\P_t']\K) \\
&-2\rho_{t+1}\E_t\left[a_{t+1}^-(s_t+\P_t'\K)1_{\{s_t+\P_t'\K> 0\}}\right](s_t+\E_t[\P_t']\K) \\
&+\gamma_t^-\left(\E_t[a_{t+1}^+(s_t+\P_t'\K)1_{\{s_t+\P_t'\K\leq 0\}}]+\E_t[a_{t+1}^-(s_t+\P_t'\K)1_{\{s_t+\P_t'\K> 0\}}]\right) \\
&+\rho_{t+1}\gamma_t^-(s_t+\E_t[\P_t']\K),
\end{align*}
with $a_{t+1}^+$, $a_{t+1}^-$, $b_{t+1}^+$ and $b_{t+1}^-$ being deterministic parameters.

The following proposition ensures that the optimizers of $\displaystyle\min_{\K\in\R^n}~F_t^+(\K)$ and $\displaystyle\min_{\K\in\R^n}~F_t^-(\K)$ are finite.

\begin{proposition}\label{prop:1}
\sl
Suppose that deterministic numbers $a_{t+1}^+$, $a_{t+1}^-$, $b_{t+1}^+$ and $b_{t+1}^-$ satisfy
$$b_{t+1}^+-(a_{t+1}^+)^2 \geq 0, \quad b_{t+1}^--(a_{t+1}^-)^2 \geq 0.$$
Then we have
\begin{align*}
\lim_{\Vert\K\Vert\rightarrow +\infty}F_t^+(\K)=+\infty, \quad \lim_{\Vert\K\Vert\rightarrow +\infty}F_t^-(\K)=+\infty,
\end{align*}
where $\Vert\K\Vert$ denotes the Euclidean norm of vector $\K$.
\end{proposition}

\noindent
\proof See Appendix A.
\endproof

According to Proposition \ref{prop:1},  we denote the finite optimizers of $\displaystyle\min_{\K\in\R^n}~F_t^+(\K)$ and $\displaystyle\min_{\K\in\R^n}~F_t^-(\K)$ as follows,
\begin{align*}
\K_t^+=\argmin_{\K\in\R^n}F_t^+(\K),\quad \K_t^-=\argmin_{\K\in\R^n}F_t^-(\K),
\end{align*}
and define the deterministic parameters $a_t^+$, $a_t^-$, $b_t^+$ and $b_t^-$, $t=0,1,\cdots,T-1$, by the following backward recursions, respectively,
\begin{align}
\nonumber a_t^+=&\,\rho_{t+1}\E_t[\P_t]\K_t^++\E_t\left[a_{t+1}^+(s_t+\P_t'\K_t^+)1_{\{s_t+\P_t'\K_t^+\geq 0\}}\right]\\
\label{eq:a+}&+\E_t\left[a_{t+1}^-(s_t+\P_t'\K_t^+)1_{\{s_t+\P_t'\K_t^+< 0\}}\right],\\
\nonumber a_t^-=&\,\rho_{t+1}\E_t[\P_t]\K_t^-+\E_t\left[a_{t+1}^+(s_t+\P_t'\K_t^-)1_{\{s_t+\P_t'\K_t^-\leq 0\}}\right] \\
\label{eq:a-}&+\E_t\left[a_{t+1}^-(s_t+\P_t'\K_t^-)1_{\{s_t+\P_t'\K_t^-> 0\}}\right],\\
\nonumber b_t^+=&\,\rho_{t+1}^2 (\K_t^+)' \E_t[\P_t\P_t']\K_t^++2\rho_{t+1}\E_t\left[a_{t+1}^+(s_t+\P_t'\K_t^+)\P_t'\K_t^+1_{\{s_t+\P_t'\K_t^+\geq 0\}}\right] \\
\nonumber &+2\rho_{t+1}\E_t\left[a_{t+1}^-(s_t+\P_t'\K_t^+)\P_t'\K_t^+1_{\{s_t+\P_t'\K_t^+< 0\}}\right]\\
\label{eq:b+}&+\E_t\left[b_{t+1}^+(s_t+\P_t'\K_t^+)^21_{\{s_t+\P_t'\K_t^+\geq 0\}}\right]+\E_t\left[b_{t+1}^-(s_t+\P_t'\K_t^+)^21_{\{s_t+\P_t'\K_t^+< 0\}}\right],\\
\nonumber b_t^-=&\,\rho_{t+1}^2 (\K_t^-)' \E_t[\P_t\P_t']\K_t^-+2\rho_{t+1}\E_t\left[a_{t+1}^+(s_t+\P_t'\K_t^-)\P_t'\K_t^-1_{\{s_t+\P_t'\K_t^-\leq 0\}}\right] \\
\nonumber &+2\rho_{t+1}\E_t\left[a_{t+1}^-(s_t+\P_t'\K_t^-)\P_t'\K_t^-1_{\{s_t+\P_t'\K_t^-> 0\}}\right]\\
\label{eq:b-}&+\E_t\left[b_{t+1}^+(s_t+\P_t'\K_t^-)^21_{\{s_t+\P_t'\K_t^-\leq 0\}}\right] +\E_t\left[b_{t+1}^-(s_t+\P_t'\K_t^-)^21_{\{s_t+\P_t'\K_t^-> 0\}}\right],
\end{align}
with terminal condition $a_t^+=a_T^- = 0$ and $b_t^+=b_T^- = 0$.

\begin{remark}\label{rem:dc}
In general, functions $F_t^+(\K)$ and $F_t^-(\K)$ are not convex functions with respect to $\K$. However, when $a_{t+1}^+\geq 0\geq a_{t+1}^-$, it is easy to show that $F_t^+(\K)$ and $F_t^-(\K)$ are d.c. functions (difference of convex functions) with respect to $\K$ (see \cite{HorstThoai1999}). In such cases, we can use the existing global search methods for d.c. functions in the literature to derive the optimizers, $\K_t^+$ and $\K_t^-$.
\end{remark}

With the above notations, we show now that the time consistent behavioral portfolio policy is a piecewise linear feedback form of the surplus or the shortage of current wealth level in the following theorem.
\begin{theorem}\label{thm:1}\sl
The time consistent behavioral portfolio policy of $(NMV_t(\gamma_t(X_t)))$ is given as follows for $t$ = 0, $\ldots$, $T-1$,
\begin{align}\label{eq:time-consistent-policy}
\u_t^{TC}=\K_t^+(X_t-\rho_t^{-1}W)1_{\{X_t\geq \rho_t^{-1}W\}}+\K_t^-(X_t-\rho_t^{-1}W)1_{\{X_t< \rho_t^{-1}W\}},
\end{align}
in which the parameters $a_t^+$, $a_t^-$, $b_t^+$ and $b_t^-$ defined in (\ref{eq:a+})-(\ref{eq:b-}) satisfy
$$b_t^+-(a_t^+)^2 \geq 0, \quad
b_t^--(a_t^-)^2 \geq 0.$$
Furthermore, the mean and the variance of the terminal wealth achieved by the time consistent behavioral portfolio policy are
\begin{align}
\label{eq:terminal-mean}\E_{0}[X_T]|_{\u^{TC}}=&\rho_0X_0+a_0^+(X_0-\rho_0^{-1}W)1_{\{X_0\geq \rho_0^{-1}W\}}+a_0^-(X_0-\rho_0^{-1}W) 1_{\{X_0< \rho_0^{-1}W\}},\\
\label{eq:terminal-var}\mbox{\rm Var}_0(X_T)|_{\u^{TC}}=& \left[(b_t^+-(a_t^+)^2)1_{\{X_0\geq \rho_0^{-1}W\}}+(b_t^--(a_t^-)^2)1_{\{X_0< \rho_0^{-1}W\}}\right](X_0-\rho_0^{-1}W)^2.
\end{align}
\end{theorem}

\noindent
\proof See Appendix B. \endproof

\begin{remark}\rm
Proposition \ref{prop:1} and Theorem \ref{thm:1} have revealed that the nested mean-variance problem $(NMV_t(\gamma_t(X_t)))$ is a well-posed problem in the sense of the existence of a finite subgame Nash equilibrium policy.
\end{remark}

\begin{remark}\rm
In our behavioral risk aversion model, the functions $F_t^+(\K)$ and $F_t^-(\K)$ are no longer convex functions with respect to $\K$. However, the optimal investment funds $\K_t^+$ and $\K_t^-$ can be derived off-line via some global search methods, thus reducing the dynamic optimization problem into $T$ static optimization problems.
\end{remark}

\begin{remark}\rm
In the proofs of Proposition \ref{prop:1} and Theorem \ref{thm:1}, the assumption of $\gamma_t^+\geq 0$ and $\gamma_t^-\geq 0$ is not used. Therefore, our main results remain valid for more general case with $\gamma_t^+\in\R$ and $\gamma_t^-\in\R$.
\end{remark}

\section{Extension to Cone Constrained Markets}
In real financial markets, realizations of ($\mathcal{F}_t$-measurable) admissible policy are often confined in a subset of $\R^n$, instead of the whole space $\R^n$. In this section, we consider the situation that the realizations of admissible policies are required to be in a cone. Such cone-type constraints have been widely adopted to model regulatory restrictions, for example, restrictions for no-short selling  (see \cite{Cui2014} and \cite{Li2001}) or non-tradeable assets. Cone-type constraints are also useful to represent portfolio restrictions, for example, the holding of the real estate stock must be no less than the bank stock. We express the feasible set of the realizations of admissible polices as $\mathcal{A}_t=\{\u_t\in\mathbb{R}^n|A \u_t\geq 0,~A\in\mathbb{R}^{m\times n}\}$ (see \cite{Cuoco1997} and \cite{Napp2003} for more examples). Now, mean-variance investors would face the following cone-constrained nested mean-variance problem,
\begin{align}
\nonumber (CNMV_t(\gamma_t(X_t))) \quad\quad \min_{\u_t} &~~\mbox{\rm Var}_t(X_T)- \gamma_t(X_t) \E_t[X_T], \\
\label{prob:MV-tc-cone} \mbox{s.t.} &~~ X_{t+1} = s_tX_t+\P_t'\u_t,\\
\nonumber &~~ X_{j+1} = s_jX_j+\P_j'\u_j^{TC}, \quad j=t+1,\cdots,T-1,\\
\nonumber &~~ \u_t \in\A_t, \\
\nonumber &~~ \u_j^{TC} \mbox{ solves } (MV_j(\gamma_j(X_j))), \quad j=t+1,\cdots,T-1,
\end{align}
with the problem in the last stage given as
\begin{align}
\nonumber (CNMV_{T-1}(\gamma_{T-1}(X_{T-1})))~~\min_{\u_{T-1}} &~~\mbox{\rm Var}_{T-1}(X_T)- \gamma_{T-1}(X_{T-1}) \E_{T-1}[X_T] , \\
\label{MV-cone} \mbox{s.t.} &~~ X_t = s_{T-1}X_{T-1}+\P_{T-1}'\u_{T-1},\\
\nonumber &~~ \u_{T-1} \in\A_T.
\end{align}

\begin{theorem}\label{thm:2}\sl
The time consistent behavioral portfolio policy of $(CNMV_t(\gamma_t(X_t)))$ is given as  follows for $t$ = 0, $\ldots$, $T-1$,
\begin{align}\label{eq:time-consistent-policy-cone}
\u_t^{TC}=\widetilde{\K}_t^+(X_t-\rho_t^{-1}W)1_{\{X_t\geq \rho_t^{-1}W\}}+\widetilde{\K}_t^-(X_t-\rho_t^{-1}W)1_{\{X_t< \rho_t^{-1}W\}}.
\end{align}
The optimal investment funds $\widetilde{\K}_t^+$ and $\widetilde{\K}_t^-$ are given by,
\begin{align*}
\widetilde{\K}_t^+=\arg\!\!\!\min_{\K\in\A_t}F_t^+(\K),\quad \widetilde{\K}_t^-=\arg\!\!\!\!\min_{\K\in-\A_t}F_t^-(\K),
\end{align*}
where $-\A_t=\{-\u_t|\u_t\in\A_t\}$ is the negative cone of $\A_t$, and the deterministic parameters in $F_t^-(\K)$ and $F_t^+(\K)$, i.e., $a_t^+$, $a_t^-$, $b_t^+$ and $b_t^-$, are computed
according to recursive functions (\ref{eq:a+})-(\ref{eq:b-}) with $\K^+$ and $\K_t^-$ replaced by $\widetilde{\K}_t^+$ and $\widetilde{\K}_t^-$, respectively.
\end{theorem}

\noindent
\proof See Appendix C. \endproof

In cone constrained markets, the time consistent behavioral portfolio policy remains a piecewise linear feedback form of the current wealth level with respect to the discounted investment target. The only difference from unconstrained markets is that we need to search the optimal investment funds in an $\A_t$ related cone instead of the entire space.

\section{Sensitivity Analysis}
In this section, we study a numerical example to analyze the property of the time consistent behavioral portfolio policy proposed in this paper.

We assume that the annual rates of return of the three risky indices follow a joint lognormal distribution. An investor with initial wealth $X_0=1$ is considering an investment opportunity of three years ($T=3$),  with his behavioral risk aversion $\gamma_t(X_t)$ expressed as follows,
\begin{align*}
\gamma_t(X_t)=\left\{\begin{array}{rl}
\gamma^+ (X_t-\rho_t^{-1}W),  & \mbox{if } X_t\geq \rho_t^{-1}W, \\
-\gamma^- (X_t-\rho_t^{-1}W), & \mbox{if } X_t< \rho_t^{-1}W.
\end{array}\right.
\end{align*}

Consider a pension fund consisting of
S\&P 500 (SP), the index of Emerging Market (EM), Small Stock (MS)
of the U.S market and a bank account with annual rate of return equal to $5\%$
($s_t=1.05$). Based on the data provided in
\cite{Elton:2007}, we list the expected values, variances and
correlation coefficients of the annual rates of return of these three indices in Table \ref{Table1}.

\begin{table}[h]
  \centering
  \caption{Data for the asset allocation example}\label{Table1}
  \begin{tabular}{lccc}
    \hline
                    & SP       & EM      & MS \\
    \hline
    Expected Return & $14\%$   & $16\% $ & $17\%$ \\
    Standard Deviation  & $18.5\%$ & $30\% $ & $24\%$ \\
    \hline
    \multicolumn{4}{c}{Correlation coefficient}\\
    \hline
    SP              & $1$      & $0.64$     & $0.79$ \\
    EM              &          & $1$        & $0.75$ \\
    MS              &          &            & $1$ \\
    \hline
  \end{tabular}
\end{table}

By simulating 20,000 sample paths for annual rates of return of the three risky indices and adopting a global search method,
we can compute the optimal investment funds $\K_t^+$, $\K_t^-$ and the deterministic parameters $a_t^+$, $a_t^-$, $b_t^+$ and $b_t^-$ backwards.  We provide the results for situations of $\gamma^+=1$ in Table \ref{Table2}. Please note that $F_t^+(\K)$ and $F_t^-(\K)$ are now d.c. functions with respect to $\K$ based on Remark \ref{rem:dc}.

\begin{table}[htb!]\small
  \centering
  \caption{Optimal investment funds and parameters ($\gamma^-\geq \gamma^+$)}\label{Table2}
  \begin{tabular}{cc|cc|cccc}
    \hline
    $\gamma^+$ & $\gamma^-$ & $\K_2^+$ & $\K_2^-$ & $a_2^+$ & $a_2^-$ & $b_2^+$ & $b_2^-$\\
    \hline
    1 & 0.5 & [0.6347,-0.0764,0.7221]' & [-0.3174,0.0382,-0.3610]' & 0.1349 & -0.0675 &  0.0857 & 0.0214 \\
    1 & 1 & [0.6347,-0.0764,0.7221]' & [-0.6347,0.0764,-0.7220]' & 0.1349 & -0.1349 &  0.0857 & 0.0857\\
    1 & 1.5 & [0.6347,-0.0764,0.7221]' & [-0.9520,0.1146,-1.0831]' & 0.1349 & -0.2024 & 0.0857 &    0.1927 \\
    1 & 2 & [0.6347,-0.0764,0.7221]' & [-1.2694,0.1528,-1.4441]' & 0.1349 & -0.2698 &   0.0857 & 0.3427\\
    1 & 2.5 & [0.6347,-0.0764,0.7221]' & [-1.5867,0.1910,-1.8051]' & 0.1349 &   -0.3373 &   0.0857 &    0.5354 \\
    \hline
        $\gamma^+$ & $\gamma^-$ & $\K_1^+$ & $\K_1^-$ & $a_1^+$ & $a_1^-$ & $b_1^+$ & $b_1^-$\\
    \hline
    1 & 0.5 & [0.4292,-0.0503,0.4775]' & [-0.3274,0.0384,-0.3641]' & 0.2492 & -0.1388 &  0.1944 &   0.0528 \\
    1 & 1 & [0.4292,-0.0503,0.4775]' & [-0.6968,0.0814,-0.7687]' & 0.2492 & -0.2759 &   0.1944 &    0.2030 \\
    1 & 1.5 & [0.4292,-0.0503,0.4775]' & [-1.0429,0.1188,-1.1317]' & 0.2492 &   -0.3996 &   0.1944 &    0.4223 \\
    1 & 2 & [0.4292,-0.0503,0.4775]' & [-1.3655,0.1550,-1.4718]' & 0.2492 & -0.5166 &   0.1944 &    0.7102 \\
    1 & 2.5 & [0.4292,-0.0503,0.4775]' & [-1.6767,0.1918,-1.8098]' & 0.2492 &   -0.6329&    0.1944 &    1.0802 \\
    \hline
        $\gamma^+$ & $\gamma^-$ & $\K_0^+$ & $\K_0^-$ & $a_0^+$ & $a_0^-$ & $b_0^+$ & $b_0^-$\\
    \hline
    1 & 0.5 & [0.3309,-0.0492,0.3432]' & [-0.3505,0.0521,-0.3635]' & 0.3505& -0.2129 &   0.3189 &   0.0948 \\
    1 & 1 & [0.3309,-0.0492,0.3432]' & [-0.7788,0.1159,-0.7974]' & 0.3505 & -0.4171&    0.3189&     0.3477 \\
    1 & 1.5 & [0.3309,-0.0492,0.3432]' & [-1.1312,0.1669,-1.1423]' & 0.3505 &   -0.5810&    0.3189&     0.6674  \\
    1 & 2 & [0.3309,-0.0492,0.3432]' & [-1.4274,0.2116,-1.4624]' & 0.3505 & -0.7289&    0.3189&     1.0643 \\
    1 & 2.5 & [0.3309,-0.0492,0.3432]' & [-1.7402,0.2516,-1.7733]' & 0.3505 &   -0.8783&    0.3189&     1.5860  \\
    \hline
\end{tabular}
\end{table}

 We can find some interesting features from Table \ref{Table2}. First, for given $\gamma^+$, the larger the value of $\gamma^-$, the larger the absolute values of $\K_t^-$, $a_t^-$ and $b_t^-$. Second, whenever the discounted investment target is less than the current wealth level (i.e., $\rho_t^{-1}W<X_t$), the investor chooses to invest a portfolio, which has almost a fixed proportion $\K_t^+$ with respect to the surplus of current wealth level. Third, when the discounted investment target is larger than the current wealth level (i.e., $\rho_t^{-1}W>X_t$), the investor with larger risk aversion coefficient $\gamma^-$ invests a larger portfolio, which has larger proportion $\K_t^-$ with respect to the shortage of the current wealth level. The third feature is quite intuitive. When $X_t>\rho_t^{-1}W$, the larger the value of $\gamma^-$, the less risk aversion of the investor at time $t$, which may result in larger risky positions.

For the situations of $\gamma^-=1$, $F_t^+(\K)$ and $F_t^-(\K)$ are also d.c. functions with respect to $\K$, and the first and third patterns remain the same as the situation with $\gamma^+=1$ (See Table \ref{Table3}). Additionally, for given $\gamma^-$, the larger the value of $\gamma^+$, the less the absolute values of $\K_t^-$, $a_t^-$ and $b_t^-$. In fact, the same pattern holds for $\gamma^+=1$. But the differences are too small to be  identified in Table \ref{Table2}.
\begin{table}[htb!]\small
  \centering
  \caption{Optimal investment portfolios and parameters ($\gamma^+\geq \gamma^-$)}\label{Table3}
  \begin{tabular}{cc|cc|cccc}
    \hline
    $\gamma^+$ & $\gamma^-$ & $\K_2^+$ & $\K_2^-$ & $a_2^+$ & $a_2^-$ & $b_2^+$ & $b_2^-$\\
    \hline
    0.5 & 1 & [0.3173,-0.0382,0.3610]' & [-0.6347,0.0764,-0.7220]' & 0.0675& -0.1349 &   0.0214 &   0.0857 \\
    1 & 1 & [0.6347,-0.0764,0.7221]' & [-0.6347,0.0764,-0.7220]' & 0.1349 & -0.1349 &  0.0857 & 0.0857\\
    1.5 & 1 & [0.9520,-0.1146,1.0831]' & [-0.6347,0.0764,-0.7220]' & 0.2024 &   -0.1349&    0.1927&     0.0857  \\
    2 & 1 & [1.2694,-0.1528,1.4441]' & [-0.6347,0.0764,-0.7220]' & 0.2698 & -0.1349&    0.3427&     0.0857 \\
    2.5 & 1 & [1.5867,-0.1910,1.8051]' & [-0.6347,0.0764,-0.7220]' & 0.3373 &   -0.1349&    0.5354&     0.0857  \\
    \hline
        $\gamma^+$ & $\gamma^-$ & $\K_1^+$ & $\K_1^-$ & $a_1^+$ & $a_1^-$ & $b_1^+$ & $b_1^-$\\
    \hline
    0.5 & 1 & [0.2526,-0.0296,0.2810]' & [-0.6985,0.0816,-0.7722]' & 0.1305& -0.2764 &   0.0508&    0.2036 \\
    1 & 1 & [0.4292,-0.0503,0.4775]' & [-0.6968,0.0814,-0.7687]' & 0.2492 & -0.2759 &   0.1944 &    0.2030\\
    1.5 & 1 & [0.5427,-0.0636,0.6037]' & [-0.6951,0.0811,-0.7653]' & 0.3562 &   -0.2755&    0.4159&     0.2024  \\
    2 & 1 & [0.6092,-0.0714,0.6777]' & [-0.6930,0.0809,-0.7615]' & 0.4533 & -0.2749&    0.7029&     0.2017 \\
    2.5 & 1 & [0.6431,-0.0754,0.7153]' & [-0.6906,0.0806,-0.7578]' & 0.5427 &   -0.2744&    1.0467&     0.2009  \\
    \hline
        $\gamma^+$ & $\gamma^-$ & $\K_0^+$ & $\K_0^-$ & $a_0^+$ & $a_0^-$ & $b_0^+$ & $b_0^-$\\
    \hline
    0.5 & 1 & [0.2149,-0.0319,0.2228]' & [-0.7838,0.1166,-0.8057]' & 0.1897& -0.4186&    0.0868&    0.3499 \\
    1 & 1 & [0.3309,-0.0492,0.3432]' & [-0.7788,0.1159,-0.7974]' & 0.3505 & -0.4171&    0.3189&     0.3477 \\
    1.5 & 1 & [0.3888,-0.0578,0.4032]' & [-0.7736,0.1155,-0.7893]' & 0.4866 &   -0.4156&    0.6575&     0.3454 \\
    2 & 1 & [0.4145,-0.0616,0.4299]' & [-0.7683,0.1148,-0.7815]' & 0.6041 & -0.4140&    1.0761&     0.3431  \\
    2.5 & 1 & [0.4226,-0.0628,0.4382]' & [-0.7638,0.1143,-0.7739]' & 0.7080 &   -0.4125&    1.5587&     0.3408  \\
    \hline
\end{tabular}
\end{table}

\begin{figure}[htb!]
\subfigure[Sharpe ratio v.s. $\gamma^-$]{
\begin{minipage}[t]{0.5\linewidth}
   \includegraphics[width=7.5cm]{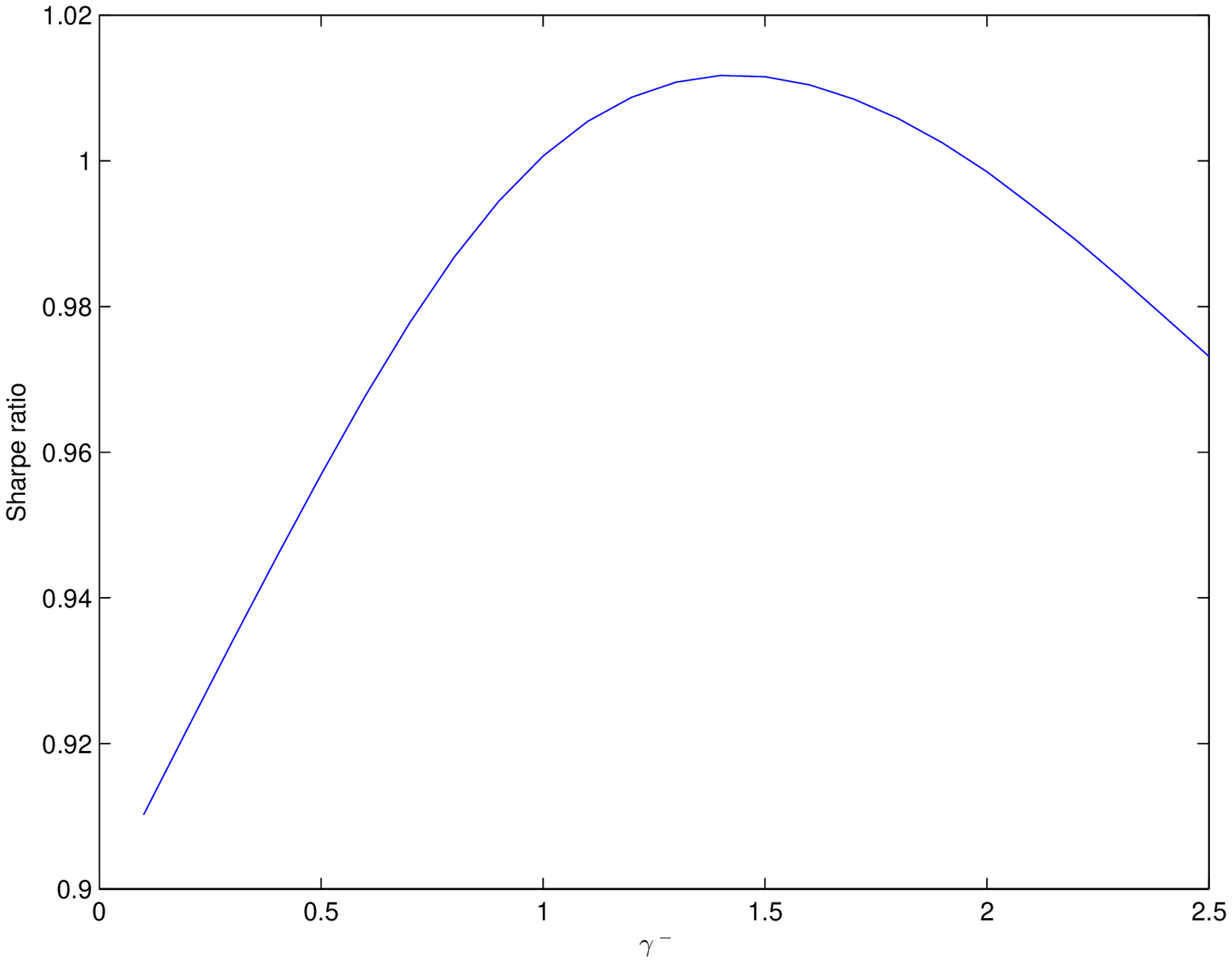}
   \label{fig:gamma-}
   \end{minipage}}%
\subfigure[Sharpe ratio v.s. $\gamma^+$]{
   \begin{minipage}[t]{0.5\linewidth}
   \includegraphics[width=7.5cm]{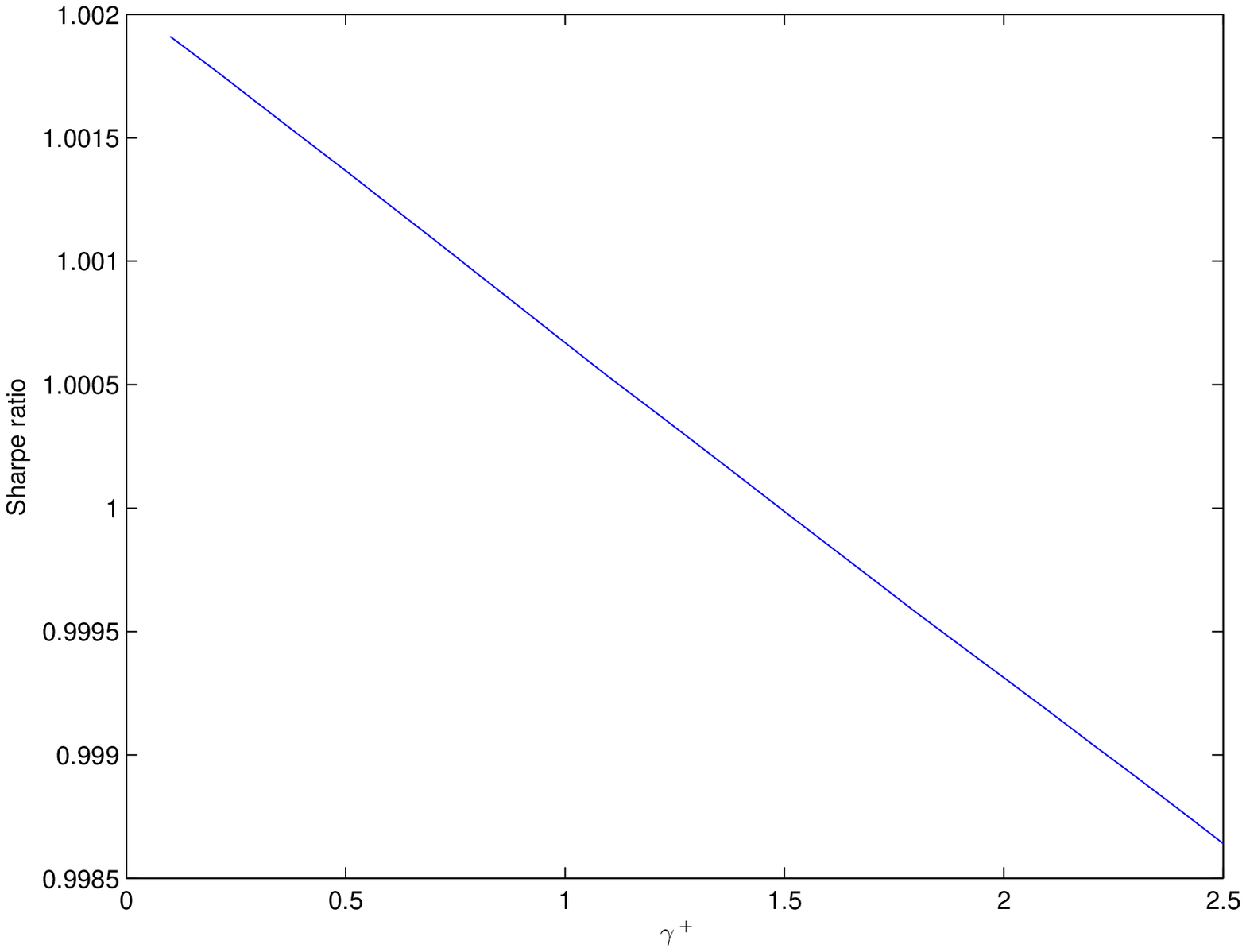}
   \label{fig:gamma+}
   \end{minipage}}%
\caption{Relationship between Sharpe ratio and parameter settings}
\end{figure}

\begin{figure}[htb!]
\subfigure[PDFs of different $\gamma^-$]{
\begin{minipage}[t]{0.5\linewidth}
   \includegraphics[width=7.5cm]{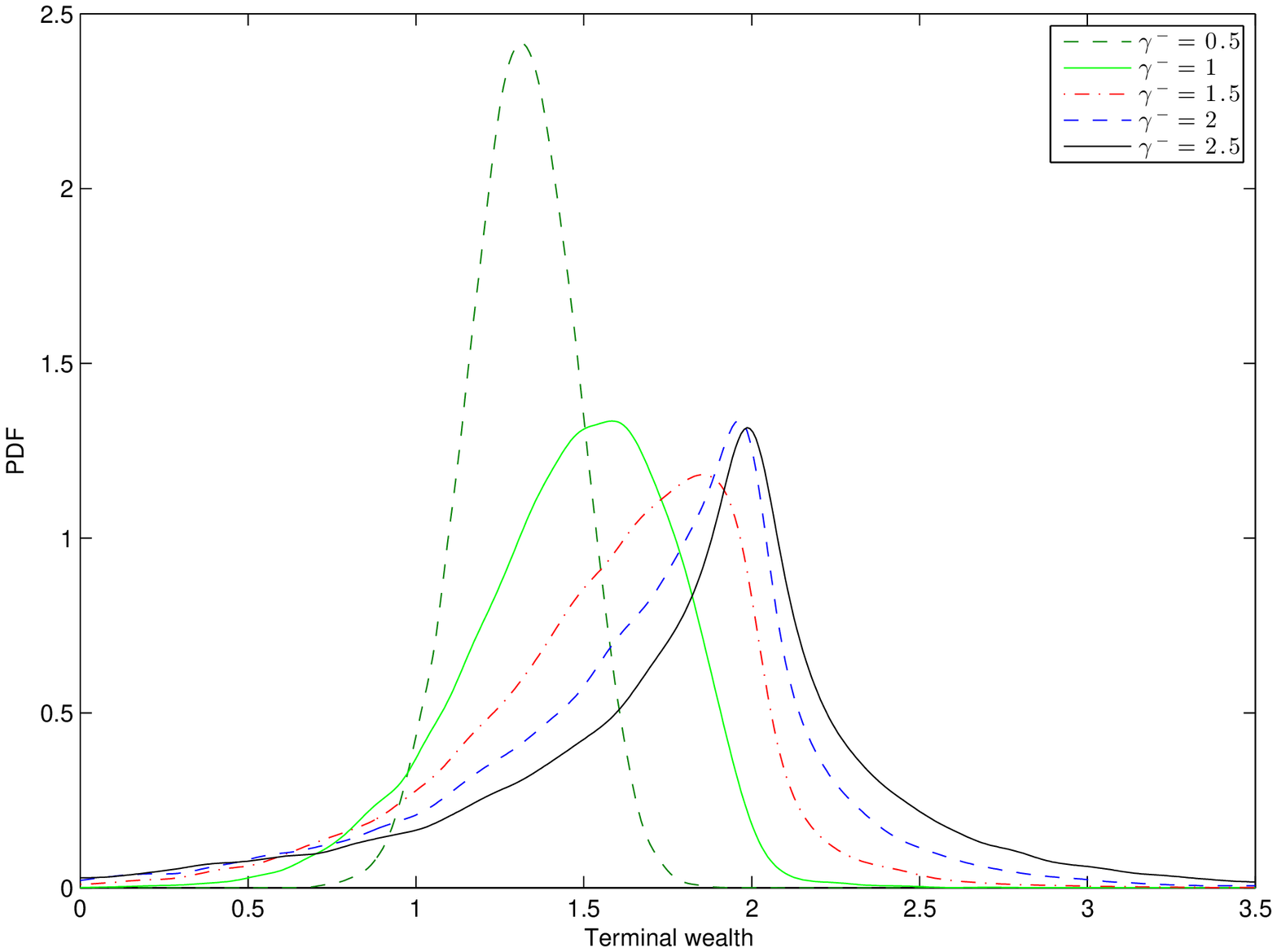}
   \label{fig:PDF-gamma-}
   \end{minipage}}%
\subfigure[PDFs of different $\gamma^+$]{
   \begin{minipage}[t]{0.5\linewidth}
   \includegraphics[width=7.5cm]{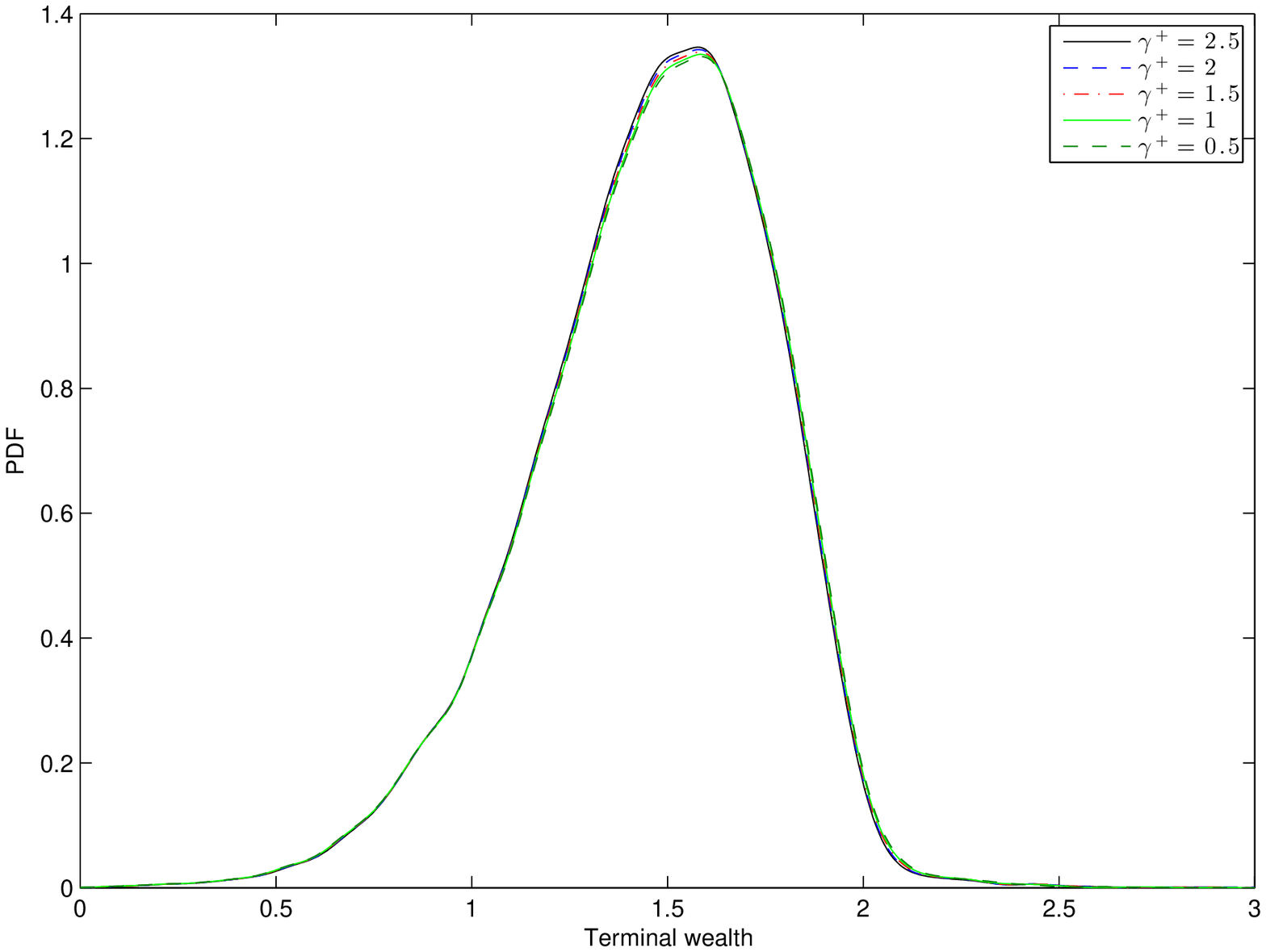}
   \label{fig:PDF-gamma+}
   \end{minipage}}%
\caption{PDFs of terminal wealth level}
\end{figure}

Next, we analyze the global investment performance of the time consistent behavioral portfolio policy proposed in this paper. We assume that all investors choose a very natural investment target $W=2$, which is twice of the value of $X_0$ and gives rise to $\rho_0^{-1}W> X_0$, i.e., the discounted investment target is no less than the initial wealth level. Figures \ref{fig:gamma-} and \ref{fig:gamma+} show the relationship of Sharpe ratio with respect to $\gamma^-$(with $\gamma^+=1$) and $\gamma^+$(with $\gamma^-=1$), respectively. Figures \ref{fig:PDF-gamma-} and \ref{fig:PDF-gamma+} show the probability density functions (PDFs) of terminal wealth levels with different risk aversion coefficients. We can see that different investors may achieve different global investment performances under their different time consistent behavioral portfolio policies. However, for the situations of $\gamma^-=1$, all the investors' time consistent policies are quite similar (see column $\K_t^-$ in Table \ref{Table3}), which results in similar Sharpe ratios and PDFs of the terminal wealth levels. In other words, under our setting, the negative risk aversion coefficient $\gamma^-$ has a higher impact on the model. The reason can be explained by the following numerical results. For the case of $\gamma^+=2.5$ and $\gamma^-=1$, it is easy to compute that
\begin{align*}
&Pr(\rho_1^{-1}W> X_1)=Pr(s_0+\P_0'\K_0^-> 0)=0.9956,\\
&Pr(\rho_2^{-1}W>X_2|\rho_1^{-1}W> X_1)=Pr(s_1+\P_1'\K_1^->0)=0.9965,\\
&Pr( \rho_3^{-1}W>X_3|\rho_2^{-1}W>X_2,)=Pr(s_2+\P_2'\K_2^->0)=0.9977,
\end{align*}
due to $\rho_0^{-1}W> X_0$. We can see that the investor has very large probability staying in the domains of $\rho_t^{-1}W> X_t$, where $\gamma^-$ is in effect.

At last, we analyze our data in a cone constrained market. We present our brief results under a no shorting constraint in Table \ref{Table4}. Due to the presence of the no-shorting constraint, the position on risky indices is forced to zero whenever the discounted investment target is larger than the current wealth level, i.e., $\K_t^-=\0$.

\begin{table}[htb!]\small
  \centering
  \caption{Optimal investment portfolios and parameters ($\gamma^+\geq \gamma^-$)}\label{Table4}
  \begin{tabular}{cc|cc|cccc}
    \hline
    $\gamma^+$ & $\gamma^-$ & $\K_2^+$ & $\K_2^-$ & $a_2^+$ & $a_2^-$ & $b_2^+$ & $b_2^-$ \\
    \hline
    1 & 0.5 & [0.6204,0,0.6594]' & [0,0,0]' & 0.1346 &    0&  0.0855&     0 \\
    1 & 1 & [0.6204,0,0.6594]' & [0,0,0]' & 0.1346 &    0&  0.0855&     0 \\
    1 & 1.5 &  [0.6204,0,0.6594]' & [0,0,0]' & 0.1346&  0&  0.0855&     0 \\
    1 & 2 &  [0.6204,0,0.6594]' & [0,0,0]' & 0.1346 &   0&  0.0855&     0 \\
    0.5 & 1 & [0.3091,0,0.3324]' & [0,0,0]' & 0.0675 &  0&  0.0215&     0 \\
    1 & 1 & [0.6204,0,0.6594]' & [0,0,0]' & 0.1346 &    0&  0.0855&     0 \\
   1.5 & 1 &  [0.9308,0,0.9897]' & [0,0,0]' &0.2020&    0&  0.1924& 0 \\
   2 & 1 & [1.2430,0,1.3160]' & [0,0,0]' &0.2690&   0&  0.3415&     0 \\
    \hline
        $\gamma^+$ & $\gamma^-$ & $\K_1^+$ & $\K_1^-$ & $a_1^+$ & $a_1^-$ & $b_1^+$ & $b_1^-$ \\
    \hline
    1 & 0.5 & [0.4189,0,0.4364]' & [0,0,0]' & 0.2484&  0&  0.1936&     0 \\
    1 & 1 & [0.4189,0,0.4364]' & [0,0,0]' & 0.2484&     0&  0.1936&     0 \\
    1 & 1.5 &  [0.4189,0,0.4364]' & [0,0,0]' & 0.2484&  0&  0.1936&     0 \\
    1 & 2 & [0.4189,0,0.4364]' & [0,0,0]' & 0.2484&     0&  0.1936&     0 \\
    0.5 & 1 & [0.2450,0,0.2565]' & [0,0,0]' &0.1302 &   0&  0.0507&     0\\
    1 & 1 & [0.4189,0,0.4364]' & [0,0,0]' & 0.2484&     0&  0.1936&     0 \\
   1.5 & 1 &  [0.5284,0,0.5516]' & [0,0,0]' &0.3550&    0&  0.4143&     0 \\
   2 & 1 & [0.5970,0,0.6193 ]' & [0,0,0]' &0.4521&  0&  0.7006&     0 \\
    \hline
        $\gamma^+$ & $\gamma^-$ & $\K_0^+$ & $\K_0^-$ & $a_0^+$ & $a_0^-$ & $b_0^+$ & $b_0^-$ \\
    \hline
    1 & 0.5 & [0.3171,0,0.3047]' & [0,0,0]' & 0.3491&  0&  0.3170&     0 \\
    1 & 1 & [0.3171,0,0.3047]' & [0,0,0]' & 0.3491&     0&  0.3170&     0 \\
    1 & 1.5 &  [0.3171,0,0.3047]' & [0,0,0]' & 0.3491&  0&  0.3170&     0 \\
    1 & 2 &  [0.3171,0,0.3047]' & [0,0,0]' & 0.3491&    0&  0.3170&     0 \\
    0.5 & 1 & [0.2071,0,0.1973]' & [0,0,0]' &0.1890& 0& 0.0864&     0 \\
    1 & 1 & [0.3171,0,0.3047]' & [0,0,0]' & 0.3491&     0&  0.3170&     0 \\
   1.5 & 1 &  [0.3747,0,0.3590]' & [0,0,0]' &0.4851&    0&  0.6547&     0 \\
   2 & 1 & [0.3999,0,0.3825]' & [0,0,0]' &0.6023&   0&  1.0720&     0 \\
    \hline
\end{tabular}
\end{table}

\section{Conclusions}

When we implement a portfolio selection methodology under a mean-risk formulation, it is crucial to assess the investor's subjective trade-off between maximizing the expected terminal wealth and minimizing the investment risk, which in turn requires good understanding of the investor's risk aversion which is in general an adaptive process of the wealth level. We propose in this paper a behavioral risk aversion model to describe the risk attitude of a mean-variance investor, which takes the piecewise linear form of the surplus or the shortage with respect to some preset investment target. Our new risk aversion model is flexible enough to incorporate the features of ``house money'' and ``breaking-even'', thus enriching the modeling power to capture the essence of the investor's risk attitude.

As the resulting dynamic mean-variance model with adaptive risk aversion is time inconsistent, we focus on its time consistent policy by solving a nested mean-variance game formulation. Fortunately, we obtain the semi-analytical time consistent behavioral portfolio policy and reveal its piecewise linear form of the surplus and the shortage with respect to the discounted wealth target. Our numerical analysis sheds light on some prominent features of the time consistent behavioral portfolio policy established in our theoretical derivations.

\section*{Appendix}
\subsection*{\bf Appendix A: The Proof of Proposition \ref{prop:1}}
\proof
Define $\xi=\Vert\K\Vert$, $\L=\K\xi^{-1}$ (which implies $\Vert\L\Vert=1$) and $y_t=\P_t'\L$. Then, for any $\L$, we have $M\geq \mbox{Var}_t(y_t)=\L'\mbox{Cov}(\P_t)\L>0$, where $M$ is the largest eigenvalue of $\mbox{Cov}(\P_t)$. 

If $y_t1_{\{y_t\geq 0\}}$ is zero, (i.e., $y_t\leq 0$ almost surely), 
we can construct an arbitrage portfolio by shorting $\L$ and holding $\L'\1$ risk-free asset. Similarly, if $y_t1_{\{y_t< 0\}}$ is zero,
(i.e., $y_t\geq 0$ almost surely), 
we also can construct an arbitrage portfolio by holding $\L$ and shorting $\L'\1$ risk-free asset. Thus, we conclude that $y_t1_{\{y_t\geq 0\}}$ and $y_t1_{\{y_t< 0\}}$ are nontrivial random variables with finite second moment.

Moreover, $\P_t$ is absolutely integrable, so do $y_t$, $y_t1_{\{y_t\geq \frac{-s_t}{\xi}\}}$ and $y_t1_{\{y_t< \frac{-s_t}{\xi}\}}$. Then, for given $\L$, we have
\begin{align*}
F_t^+(\K)=\tilde{F}_t^+(\xi),
\end{align*}
where
\begin{align*}
\tilde{F}_t^+(\xi)
=& \, \rho_{t+1}^2\mbox{Var}_t(y_t)\xi^2+\E_t\left[(2\rho_{t+1}a_{t+1}^+ + b_{t+1}^+)(s_t+\xi y_t)^21_{\{y_t\geq \frac{-s_t}{\xi}\}}\right] \\
&+\E_t\left[(2\rho_{t+1}a_{t+1}^- + b_{t+1}^-)(s_t+\xi y_t)^21_{\{y_t< \frac{-s_t}{\xi}\}}\right] \\
&-\left(\E_t\left[a_{t+1}^+(s_t+\xi y_t)1_{\{y_t\geq \frac{-s_t}{\xi}\}}\right] +\E_t\left[a_{t+1}^-(s_t+\xi y_t)1_{\{y_t< \frac{-s_t}{\xi}\}}\right]\right)^2\\
&-2\rho_{t+1} \left(\E_t\left[a_{t+1}^+(s_t+\xi y_t)1_{\{y_t\geq \frac{-s_t}{\xi}\}}\right] +\E_t\left[a_{t+1}^-(s_t+\xi y_t)1_{\{y_t< \frac{-s_t}{\xi}\}}\right]\right)(s_t+\E_t[y_t]\xi)\\
& -\gamma_t^+\left(\E_t\left[a_{t+1}^+(s_t+\xi y_t)1_{\{y_t\geq \frac{-s_t}{\xi}\}}\right]+\E_t\left[a_{t+1}^-(s_t+\xi y_t)1_{\{y_t<\frac{-s_t}{\xi}\}}\right]\right)\\
&-\rho_{t+1}\gamma_t^+(s_t+\E_t[y_t]\xi).
\end{align*}
Furthermore, we have
\begin{align*}
\tilde{F}_t^+(\xi)
\geq & \,\rho_{t+1}^2\mbox{Var}_t(y_t)\xi^2+(a_{t+1}^+)^2\E_t\left[y_t^21_{\{y_t\geq \frac{-s_t}{\xi}\}}\right]\xi^2+(a_{t+1}^-)^2\E_t\left[ y_t^21_{\{y_t< \frac{-s_t}{\xi}\}}\right]\xi^2\\
&-\left(a_{t+1}^+\E_t\left[y_t1_{\{y_t\geq \frac{-s_t}{\xi}\}}\right]\xi +a_{t+1}^-\E_t\left[ y_t1_{\{y_t< \frac{-s_t}{\xi}\}}\right]\xi\right)^2\\
&+2\rho_{t+1}\left(a_{t+1}^+\E_t\left[y_t^21_{\{y_t\geq \frac{-s_t}{\xi}\}}\right]\xi^2+a_{t+1}^-\E_t\left[y_t^21_{\{y_t< \frac{-s_t}{\xi}\}}\right]\xi^2\right)\\
&-2\rho_{t+1}\left(a_{t+1}^+\E_t\left[y_t1_{\{y_t\geq \frac{-s_t}{\xi}\}}\right]+a_{t+1}^-\E_t\left[y_t1_{\{y_t< \frac{-s_t}{\xi}\}}\right]\right)\E_t[y_t]\xi^2+O(\xi)\\
=& \, \rho_{t+1}^2\mbox{Var}_t\left(y_t1_{\{y_t\geq \frac{-s_t}{\xi}\}}\right)\xi^2+\rho_{t+1}^2\mbox{Var}_t\left( y_t1_{\{y_t< \frac{-s_t}{\xi}\}}\right)\xi^2\\
&+2\rho_{t+1}^2\left(\E_t\left[y_t1_{\{y_t\geq \frac{-s_t}{\xi}\}} y_t1_{\{y_t< \frac{-s_t}{\xi}\}}\right]-\E_t\left[y_t1_{\{y_t\geq \frac{-s_t}{\xi}\}}\right]\E_t\left[ y_t1_{\{y_t< \frac{-s_t}{\xi}\}}\right]\right)\xi^2\\
&+(a_{t+1}^+)^2\mbox{Var}_t\left(y_t1_{\{y_t\geq \frac{-s_t}{\xi}\}}\right)\xi^2+(a_{t+1}^-)^2\mbox{Var}_t\left( y_t1_{\{y_t< \frac{-s_t}{\xi}\}}\right)\xi^2\\
&+2a_{t+1}^+a_{t+1}^-\left(\E_t\left[y_t1_{\{y_t\geq \frac{-s_t}{\xi}\}} y_t1_{\{y_t< \frac{-s_t}{\xi}\}}\right]-\E_t\left[y_t1_{\{y_t\geq \frac{-s_t}{\xi}\}}\right]\E_t\left[ y_t1_{\{y_t< \frac{-s_t}{\xi}\}}\right]\right)\xi^2\\
&+2\rho_{t+1}\left(a_{t+1}^+\mbox{Var}_t\left(y_t1_{\{y_t\geq \frac{-s_t}{\xi}\}}\right)\xi^2+a_{t+1}^-\mbox{Var}_t\left( y_t1_{\{y_t< \frac{-s_t}{\xi}\}}\right)\xi^2\right)\\
&+2\rho_{t+1}a_{t+1}^+\left(\E_t\left[y_t1_{\{y_t\geq \frac{-s_t}{\xi}\}} y_t1_{\{y_t< \frac{-s_t}{\xi}\}}\right]-\E_t\left[y_t1_{\{y_t\geq \frac{-s_t}{\xi}\}}\right]\E_t\left[ y_t1_{\{y_t< \frac{-s_t}{\xi}\}}\right]\right)\xi^2\\
&+2\rho_{t+1}a_{t+1}^-\left(\E_t\left[y_t1_{\{y_t\geq \frac{-s_t}{\xi}\}} y_t1_{\{y_t< \frac{-s_t}{\xi}\}}\right]-\E_t\left[y_t1_{\{y_t\geq \frac{-s_t}{\xi}\}}\right]\E_t\left[ y_t1_{\{y_t< \frac{-s_t}{\xi}\}}\right]\right)\xi^2 \\
&+O(\xi)\\
=&\left[\rho_{t+1}+a_{t+1}^+,\rho_{t+1}+a_{t+1}^-\right]
\mbox{Cov}_t\left[\begin{array}{cc}
y_t1_{\{y_t\geq \frac{-s_t}{\xi}\}}\\
y_t1_{\{y_t< \frac{-s_t}{\xi}\}}
 \end{array}\right]
 \left[\begin{array}{cc}
\rho_{t+1}+a_{t+1}^+\\ \rho_{t+1}+a_{t+1}^-
 \end{array}\right]\xi^2 +O(\xi),
\end{align*}
where $O(\xi)$ is the infinity of the same order as $\xi$ and the second equality holds due to the fact of $\E_t\left[y_t1_{\{y_t\geq \frac{-s_t}{\xi}\}} y_t1_{\{y_t< \frac{-s_t}{\xi}\}}\right]=0$. Hence,
\begin{align*}
&\lim_{\xi\rightarrow +\infty}\tilde{F}_t^+(\xi)=\lim_{\xi\rightarrow+\infty}\left[\rho_{t+1}+a_{t+1}^+,\rho_{t+1}+a_{t+1}^-\right]
\mbox{Cov}_t\left[\begin{array}{cc}
y_t1_{\{y_t\geq 0\}}\\
y_t1_{\{y_t< 0\}}
 \end{array}\right]
 \left[\begin{array}{cc}
\rho_{t+1}+a_{t+1}^+\\ \rho_{t+1}+a_{t+1}^-
 \end{array}\right]\xi^2 \\
& \quad\quad\quad\quad\quad\quad\quad +O(\xi)=+\infty.
\end{align*}
Based on the discussion for all possible $\L$, we make our conclusion for $F_t^+(\K)$. Similarly we can prove the result of $F_t^-(\K)$.
\endproof

\subsection*{\bf Appendix B: The Proof of Theorem \ref{thm:1}}
\proof
Let $Y_t = X_t - \rho_t^{-1}W$. Then,
\begin{align*}
Y_{t+1} & = X_{t+1} - \rho_{t+1}^{-1}W \\
& = s_tX_t + \P_t'\u_t - \rho_{t+1}^{-1}W \\
& = s_t(X_t - \rho_t^{-1}W) + \P_t'\u_t \\
& = s_tY_t + \P_t'\u_t,
\end{align*}
and $\gamma_t(X_t)$ can be re-written into
\begin{align*}
\gamma_t(X_t) = \hat\gamma_t(Y_t)=\left\{\begin{array}{rl}
\gamma_t^+ Y_t,  & \mbox{if } Y_t \geq 0, \\
-\gamma_t^- Y_t, & \mbox{if } Y_t < 0.
\end{array}\right.
\end{align*}
Also, we have $\mbox{\rm Var}_t(X_T)=\mbox{Var}_t(Y_T)$ according to the variance property.
Hence, problem $(MV_t(\gamma_t(X_t)))$ in (\ref{prob:MV_t}) can be equivalently reduced into the following problem
\begin{align}
\nonumber \quad\quad \min &~~\mbox{\rm Var}_t(Y_T)- \hat\gamma_t(Y_t) \E_t[Y_T] - \hat\gamma_t(Y_t)W, \\
\label{prob:MV_t_Y} \mbox{s.t.} &~~ Y_{j+1} = s_jY_j+\P_j'\u_j, \quad  j=t,t+1,\cdots,T-1,
\end{align}
where $\mbox{\rm Var}_t(Y_T)=\mbox{Var}(Y_T|Y_t)$ and $\E_t[Y_T]=\E[Y_T|Y_t]$.

At time $t$ ($t=0,1,\cdots, T$), the investor faces the following optimization problem,
\begin{align}\label{eq:cost-to-go}
\min_{\u_t}~~J_t(Y_t;\u_t) =\Big(\E_t[Y_T^2] - (\E_t[Y_T])^2\Big) - \hat\gamma_t(Y_t)\E_t[Y_T] - \hat\gamma_t(Y_t)W,
\end{align}
where the conditional expectations $\E_t[Y_T]=\E[Y_T|Y_t]$ and $\E_t[Y_T^2]=\E[Y_T^2|Y_t]$ are computed along the policy $\{\u_t,\u_{t+1}^{TC},\cdots, \u_{T-1}^{TC}\}$.

We now prove by induction that the following two expressions,
\begin{align}
\label{eq:conditional-mean}
  \E_t[Y_T] = & \rho_tY_t+a_t^+Y_t1_{\{Y_t \geq 0\}}+a_t^-Y_t1_{\{Y_t < 0\}}, \\
\label{eq:conditional-mean-2}
\E_t[Y_T^2] = & \rho_t^2Y_t^2+(2\rho_ta_t^+ + b_t^+)Y_t^21_{\{Y_t \geq 0\}}+(2\rho_ta_t^- + b_t^-)Y_t^21_{\{Y_t < 0\}},
\end{align}
hold along the time consistent policy, $\{\u_t^{TC},\u_{t+1}^{TC},\cdots, \u_{T-1}^{TC}\}$, at time $t$.

At time $T$, we have
\begin{align*}
\E_T[Y_T]=Y_T, \quad \E_T[Y_T^2]=Y_T^2,
\end{align*}
with  $a_t^+=a_T^- = 0$ and $b_t^+=b_T^- = 0$. Assume that expressions of the first moment and the second moment in (\ref{eq:conditional-mean}) and (\ref{eq:conditional-mean-2}), respectively, hold at time $t+1$ along the time consistent policy $\{\u_{t+1}^{TC},\cdots, \u_{T-1}^{TC}\}$. We will prove that these two expressions still hold at time $t$ and the corresponding time consistent policy is given by (\ref{eq:time-consistent-policy}).

As the dynamic of period $t$ is
\begin{align*}
Y_{t+1}=s_tY_t+\P_t'\u_t.
\end{align*}
It follows from the policy $\{\u_t,\u_{t+1}^{TC},\cdots, \u_{T-1}^{TC}\}$ that we have
\begin{align}
\nonumber & \E_t[Y_T] = \E_t\big[\E_{t+1}[Y_T]\big] \\
\nonumber & = \E_t\!\!\left[\rho_{t+1}Y_{t+1}+a_{t+1}^+Y_{t+1}1_{\{Y_{t+1} \geq 0\}}+a_{t+1}^-Y_{t+1}1_{\{Y_{t+1} < 0\}}\right] \\
\nonumber & = \E_t[\rho_{t+1}(s_tY_t+\P_t'\u_t)]+\E_t\left[a_{t+1}^+(s_tY_t+\P_t'\u_t)1_{\{s_tY_t+\P_t'\u_t \geq 0\}}\right] \\
 & \;\;\; +\E_t\!\!\left[a_{t+1}^-(s_tY_t+\P_t'\u_t)1_{\{s_tY_t+\P_t'\u_t < 0\}}\right] \label{eq:expect}
\end{align}
and
\begin{align}
\nonumber & \E_t[Y_T^2] = \E_t\big[\E_{t+1}[Y_T^2]\big] \\
\nonumber & = \E_t\!\Big[\rho_{t+1}^2Y_{t+1}^2+(2\rho_{t+1}a_{t+1}^+ + b_{t+1}^+)Y_{t+1}^21_{\{Y_{t+1} \geq 0\}} \\
\nonumber & \;\;\; + (2\rho_{t+1}a_{t+1}^- + b_{t+1}^-)Y_{t+1}^21_{\{Y_{t+1} < 0\}}\Big] \\
\nonumber & = \E_t[\rho_{t+1}^2(s_tY_t+\P_t'\u_t)^2]\\
\nonumber & \;\;\; +\E_t\!\!\left[(2\rho_{t+1}a_{t+1}^+ + b_{t+1}^+)(s_tY_t+\P_t'\u_t)^21_{\{s_tY_t+\P_t'\u_t \geq 0\}}\right] \\
\label{eq:expect-2} & \;\;\; +\E_t\!\!\left[(2\rho_{t+1}a_{t+1}^- + b_{t+1}^-)(s_tY_t+\P_t'\u_t)^21_{\{s_tY_t+\P_t'\u_t < 0\}}\right].
\end{align}
For $Y_t > 0$, we denote any admissible policy as $\u_t=\K Y_t$ with $\K\in \R^n$. Then the cost functional can be expressed as
\begin{align*}
&J_t(Y_t;\u_t) = \Big(\E_t[Y_T^2] - (\E_t[Y_T])^2\Big) - \gamma_t^+Y_t\E_t[Y_T] - \gamma_t^+Y_tW \\
& = \E_t[\rho_{t+1}^2(s_tY_t+\P_t'\u_t)^2]- \big(\E_t[\rho_{t+1}(s_tY_t+\P_t'\u_t)]\big)^2 \\
& \;\;\; +\E_t\!\!\left[(2\rho_{t+1}a_{t+1}^+ + b_{t+1}^+)(s_tY_t+\P_t'\u_t)^21_{\{s_tY_t+\P_t'\u_t \geq 0\}}\right] \\
& \;\;\; +\E_t\!\!\left[(2\rho_{t+1}a_{t+1}^- + b_{t+1}^-)(s_tY_t+\P_t'\u_t)^21_{\{s_tY_t+\P_t'\u_t < 0\}}\right] \\
& \;\;\; -\left(\E_t\!\!\left[a_{t+1}^+(s_tY_t+\P_t'\u_t)1_{\{s_tY_t+\P_t'\u_t\geq 0\}}\right]+\E_t\!\!\left[a_{t+1}^-(s_tY_t+\P_t'\u_t)1_{\{s_tY_t+\P_t'\u_t< 0\}}\right]\right)^2 \\
& \;\;\; -2\rho_{t+1}\Big(\!\E_t\!\!\left[a_{t+1}^+(s_tY_t+\P_t'\u_t)1_{\{s_tY_t+\P_t'\u_t \geq 0\}}\right] \\
& \;\;\; +\E_t\!\!\left[a_{t+1}^-(s_tY_t+\P_t'\u_t)1_{\{s_tY_t+\P_t'\u_t< 0\}}\right]\!\!\Big)\E_t[s_tY_t+\P_t'\u_t] \\
& \;\;\; -\gamma_t^+Y_t\Big(\!\E_t\left[a_{t+1}^+(s_tY_t+\P_t'\u_t)1_{\{s_tY_t+\P_t'\u_t \geq 0\}}\right]+\E_t\!\!\left[a_{t+1}^-(s_tY_t+\P_t'\u_t)1_{\{s_tY_t+\P_t'\u_t< 0\}}\right]\!\!\Big) \\
& \;\;\; -\rho_{t+1}\gamma_t^+Y_t\E_t[s_tY_t+\P_t'\u_t]-\gamma_t^+Y_tW \\
& = Y_t^2 \Big\{\rho_{t+1}^2 \K' (\E_t[\P_t\P_t']-\E_t[\P_t']\E_t[\P_t])\K \\
& \;\;\; +\E_t\!\!\left[(2\rho_{t+1}a_{t+1}^+ + b_{t+1}^+)(s_t+\P_t'\K)^21_{\{s_t+\P_t'\K\geq 0\}}\right] \\
& \;\;\; +\E_t\!\!\left[(2\rho_{t+1}a_{t+1}^- + b_{t+1}^-)(s_t+\P_t'\K)^21_{\{s_t+\P_t'\K< 0\}}\right] \\
& \;\;\; -\left(\E_t\!\!\left[a_{t+1}^+(s_t+\P_t'\K)1_{\{s_t+\P_t'\K\geq 0\}}\right] +\E_t\left[a_{t+1}^-(s_t+\P_t'\K)1_{\{s_t+\P_t'\K< 0\}}\right]\right)^2 \\
& \;\;\; -2\rho_{t+1} \E_t\!\!\left[a_{t+1}^+(s_t+\P_t'\K)1_{\{s_t+\P_t'\K\geq 0\}}\right](s_t+\E_t[\P_t']\K) \\
& \;\;\; -2\rho_{t+1}\E_t\!\!\left[a_{t+1}^-(s_t+\P_t'\K)1_{\{s_t+\P_t'\K< 0\}}\right](s_t+\E_t[\P_t']\K) \\
& \;\;\; -\gamma_t^+\left(\E_t\!\!\left[a_{t+1}^+(s_t+\P_t'\K)1_{\{s_t+\P_t'\K\geq 0\}}\right] +\E_t\left[a_{t+1}^-(s_t+\P_t'\K)1_{\{s_t+\P_t'\K< 0\}}\right]\right) \\
& \;\;\; -\rho_{t+1}\gamma_t^+(s_t+\E_t[\P_t']\K)\Big\}-\gamma_t^+Y_tW \\
& = Y_t^2  F_t^+(\K)-\gamma_t^+Y_tW.
\end{align*}
Applying Proposition \ref{prop:1} yields the optimal time consistent policy at time $t$,
\begin{align*}
\u_t^{TC}=\argmin_{\u_t\in\R^n} J_t(Y_t;\u_t)=\K_t^+Y_t.
\end{align*}
Then, substituting the above optimal time consistent policy back into (\ref{eq:expect}) and (\ref{eq:expect-2}) gives rise to
\begin{align*}
\E_t[Y_T]
=& \,\rho_tY_t+Y_t\Big(\rho_{t+1}\E_t[\P_t]\K_t^++\E_t\!\!\left[a_{t+1}^+(s_t+\P_t'\K_t^+)1_{\{s_t+\P_t'\K_t^+ \geq 0\}}\right] \\
&+\E_t\!\!\left[a_{t+1}^-(s_t+\P_t'\K_t^+)1_{\{s_t+\P_t'\K_t^+ < 0\}}\right]\Big) \\
=&\,\rho_tY_t+a_t^+Y_t
\end{align*}
and
\begin{align*}
\E_t[Y_T^2] = & \, \rho_t^2Y_t^2+2\rho_tY_t^2\Big(\rho_{t+1}\E_t[\P_t]\K_t^++\E_t\left[a_{t+1}^+(s_t+\P_t'\K_t^+)1_{\{s_t+\P_t'\K_t^+ \geq 0\}}\right] \\
&+\E_t\left[a_{t+1}^-(s_t+\P_t'\K_t^+)1_{\{s_t+\P_t'\K_t^+< 0\}}\right]\Big) \\
&+\Big(\rho_{t+1}^2 (\K_t^+)' \E_t[\P_t\P_t']\K_t^++2\rho_{t+1}\E_t\left[a_{t+1}^+(s_t+\P_t'\K_t^+)\P_t'\K_t^+1_{\{s_t+\P_t'\K_t^+\geq 0\}}\right] \\
&+2\rho_{t+1}\E_t\left[a_{t+1}^-(s_t+\P_t'\K_t^+)\P_t'\K_t^+1_{\{s_t+\P_t'\K_t^+ < 0\}}\right] \\
&+\E_t\left[b_{t+1}^+(s_t+\P_t'\K_t^+)^21_{\{s_t+\P_t'\K_t^+\geq 0\}}\right]+\E_t\left[b_{t+1}^-(s_t+\P_t'\K_t^+)^21_{\{s_t+\P_t'\K_t^+< 0\}}\right]\Big)Y_t^2 \\
=&\,\rho_t^2Y_t^2+(2\rho_ta_t^++b_t^+)Y_t^2.
\end{align*}
Furthermore,
\begin{align*}
\mbox{Var}_t(Y_T)=\E_t[Y_T^2]-(\E_t[Y_T])^2
=(b_t^+ - ( a_t^+)^2)Y_t^2
\geq 0,
\end{align*}
implies $b_t^+ - ( a_t^+)^2\geq 0$.

For $Y_t < 0$, we denote any admissible policy as $\u_t=\K Y_t$ with $\K\in\R^n$. Then the cost functional can be expressed as
\begin{align*}
J_t(Y_t;\u_t)
= & \, Y_t^2 \Big\{\rho_{t+1}^2 \K' (\E_t[\P_t\P_t']-\E_t[\P_t']\E_t[\P_t])\K \\
&+\E_t\left[(2\rho_{t+1}a_{t+1}^+ + b_{t+1}^+)(s_t+\P_t'\K)^21_{\{s_t+\P_t'\K\leq 0\}}\right] \\
&+\E_t\left[(2\rho_{t+1}a_{t+1}^- + b_{t+1}^-)(s_t+\P_t'\K)^21_{\{s_t+\P_t'\K> 0\}}\right] \\
&-\left(\E_t\left[a_{t+1}^+(s_t+\P_t'\K)1_{\{s_t+\P_t'\K\leq 0\}}\right] +\E_t\left[a_{t+1}^-(s_t+\P_t'\K)1_{\{s_t+\P_t'\K> 0\}}\right]\right)^2 \\
&-2\rho_{t+1} \E_t\left[a_{t+1}^+(s_t+\P_t'\K)1_{\{s_t+\P_t'\K\leq 0\}}\right] (s_t+\E_t[\P_t']\K) \\
&-2\rho_{t+1}\E_t\left[a_{t+1}^-(s_t+\P_t'\K)1_{\{s_t+\P_t'\K> 0\}}\right](s_t+\E_t[\P_t']\K) \\
&+\gamma_t^-\left(\E_t\left[a_{t+1}^+(s_t+\P_t'\K)1_{\{s_t+\P_t'\K\leq 0\}}\right] +\E_t\left[a_{t+1}^-(s_t+\P_t'\K)1_{\{s_t+\P_t'\K> 0\}}\right]\right) \\
&+\rho_{t+1}\gamma_t^-(s_t+\E_t[\P_t']\K)\Big\}+\gamma_t^-Y_tW \\
= &\,Y_t^2  F_t^-(\K)+\gamma_t^-Y_tW.
\end{align*}
Applying Proposition \ref{prop:1} yields the optimal time consistent policy at time $t$,
\begin{align*}
\u_t^{TC}=\argmin_{\u_t\in\R^n} J_t(Y_t;\u_t)=\K_t^-Y_t.
\end{align*}
Then, substituting the above optimal time consistent policy back into (\ref{eq:expect}) and (\ref{eq:expect-2}) gives rise to
\begin{align*}
\E_t[Y_T]
=&\,\rho_tY_t+Y_t\Big(\rho_{t+1}\E_t[\P_t]\K_t^-+\E_t\left[a_{t+1}^+(s_t+\P_t'\K_t^-)1_{\{s_t+\P_t'\K_t^-\leq 0\}}\right] \\
&+\E_t\left[a_{t+1}^-(s_t+\P_t'\K_t^-)1_{\{s_t+\P_t'\K_t^-> 0\}}\right]\Big) \\
=&\,\rho_tY_t+a_t^-Y_t
\end{align*}
and
\begin{align*}
\E_t[Y_T^2] = & \, \rho_t^2Y_t^2+2\rho_tY_t^2\Big(\rho_{t+1}\E_t[\P_t]\K_t^-+\E_t\left[a_{t+1}^+(s_t+\P_t'\K_t^-)1_{\{s_t+\P_t'\K_t^-\leq 0\}}\right] \\
&+\E_t\left[a_{t+1}^-(s_t+\P_t'\K_t^-)1_{\{s_t+\P_t'\K_t^-> 0\}}\right]\Big) \\
&+\Big(\rho_{t+1}^2 (\K_t^-)' \E_t[\P_t\P_t']\K_t^-+2\rho_{t+1}\E_t\left[a_{t+1}^+(s_t+\P_t'\K_t^-)\P_t'\K_t^-1_{\{s_t+\P_t'\K_t^-\leq 0\}}\right] \\
&+2\rho_{t+1}\E_t\left[a_{t+1}^-(s_t+\P_t'\K_t^-)\P_t'\K_t^-1_{\{s_t+\P_t'\K_t^-> 0\}}\right] \\
& +\E_t\left[b_{t+1}^+(s_t+\P_t'\K_t^-)^21_{\{s_t+\P_t'\K_t^-\leq 0\}}\right]+\E_t\left[b_{t+1}^-(s_t+\P_t'\K_t^-)^21_{\{s_t+\P_t'\K_t^-> 0\}}\right]\Big)Y_t^2 \\
= & \,\rho_t^2Y_t^2+(2\rho_ta_t^-+b_t^-)Y_t^2.
\end{align*}
Furthermore,
\begin{align*}
\mbox{Var}_t(Y_T)=\E_t[Y_T^2]-(\E_t[Y_T])^2
=(b_t^- - ( a_t^-)^2)Y_t^2
\geq0,
\end{align*}
implies $b_t^- - ( a_t^-)^2\geq 0$.

For $Y_t=0$, the cost functional reduces to the conditional variance of the terminal wealth along policy $\{\u_t,\u_{t+1}^{TC},\cdots, \u_{T-1}^{TC}\}$, which can be expressed as
\begin{align*}
J_t(Y_t;\u_t)
= & \, \rho_{t+1}^2 \u_t' (\E_t[\P_t\P_t']-\E_t[\P_t']\E_t[\P_t])\u_t\\
&+\E_t\left[b_{t+1}^+(\P_t'\u_t)^21_{\{\P_t'\u_t\geq 0\}}\right]+\E_t\left[b_{t+1}^-(\P_t'\u_t)^21_{\{\P_t'\u_t< 0\}}\right]\\
&-\Big(\E_t\left[a_{t+1}^+\P_t'\u_t1_{\{\P_t'\u_t\geq 0\}}\right] +\E_t\left[a_{t+1}^-\P_t'\u_t1_{\{\P_t'\u_t< 0\}}\right]\Big)^2\\
&+ 2\rho_{t+1} \left( \E_t\left[a_{t+1}^+(\P_t'\u_t)^21_{\{\P_t'\u_t\geq 0\}}\right]+ \E_t\left[a_{t+1}^-(\P_t'\u_t)^21_{\{\P_t'\u_t< 0\}}\right]\right)\\
&-2\rho_{t+1}\Big(\E_t\left[a_{t+1}^+\P_t'\u_t1_{\{\P_t'\u_t\geq 0\}}\right] +\E_t\left[a_{t+1}^-\P_t'\u_t1_{\{\P_t'\u_t< 0\}}\right]\Big) \E_t[\P_t']\u_t\\
\geq \, & 0.
\end{align*}
It is not difficult to conclude that $\u_t^{TC}=\argmax_{\u_t\in\R^n}J_t(Y_t;\u_t)=\0$.

Therefore, along the time consistent policy $\{\u_t^{TC},\u_{t+1}^{TC},\cdots, \u_{T-1}^{TC}\}$, expressions (\ref{eq:conditional-mean}) and (\ref{eq:conditional-mean-2}) hold at time $t$, which completes our proof.
\endproof

\subsection*{\bf Appendix C: The Proof of Theorem \ref{thm:2}}
\proof Following the technique in the proof of Theorem \ref{thm:1}, we can derive the main results directly with the following specifics.

i) For $X_t>\rho_t^{-1}W$, we denote any admissible policy as $\u_t=\K (X_t-\rho_t^{-1}W)$ with $\K\in \A_t$.

ii) For $X_t<\rho_t^{-1}W$, we denote any admissible policy as $\u_t=\K (X_t-\rho_t^{-1}W)$ with $\K\in-\A_t$, where $-\A_t$ is the negative cone of $\A_t$.

iii) For $X_t=\rho_t^{-1}W$, we can similarly prove $\u_t^{TC}=\0$.

Therefore, we have
\begin{align*}
\widetilde{\K}_t^+=\argmin_{\K\in\A_t}F_t^+(\K),\quad \widetilde{\K}_t^-=\argmin_{\K\in-\A_t}F_t^-(\K).
\end{align*}
This completes the proof.
\endproof

\end{document}